\documentstyle[aps,preprint,tighten]{revtex}
\draft
\begin{document}

\author{Wei-Min Sun\footnote{e-mail address: sunwm@chenwang.nju.edu.cn}
Xiang-Song Chen and Fan Wang\footnote{e-mail address:
fgwang@chenwang.nju.edu.cn}} 
\address{Department of Physics and Center for
Theoretical Physics\\  Nanjing University, Nanjing 210093, China} \title{A
Note on Invariant Measure on the Local Gauge Group} \maketitle

\begin{abstract}
In this paper we investigate the problem of the existence of invariant measures
on the local gauge group. We prove that it is impossible to define a
{\it finite} translationally invariant measure on the local gauge group
$C^{\infty}({\bf R}^n, G)$(where $G$ is an arbitrary matrix Lie group).
\end{abstract}

Functional integral over the local gauge group is a commonly employed
technique in the study of gauge field theories. Its earliest appearance
in the literature is Faddeev and Popov's paper in 1967 \cite{Faddeev}, 
which investigated the quantization of non-Abelian gauge fields. In \cite{Huang}
this technique was used to construct a projection operator which implements
the Gauss law, and in \cite{Kogan} it was used to construct a gauge invariant
ground state wave-functional. In all these applications a key point is
the assumption that we can define a translationally invariant measure on
the local gauge group. It is known that on locally compact groups 
this is possible \cite{Halmos}. On non-locally-compact 
groups we cannot guarantee such a measure exists. In the case
of vector spaces a standard result \cite{Yamasaki} says that an invariant measure
exists only when the vector space is finite dimensional, in which case it
is locally compact. The local gauge group is infinite dimensional and
not locally compact. Since functional integral over the local gauge
group is an important technique the existence of an invariant measure 
deserves a careful study. 
In \cite{SCW} it is proved that on the local gauge group
$C^{\infty}({\bf R}^n,G)$ there exists no finite translationally invariant Borel measure(Haar measure). 
In this paper we study the case of the local gauge group $C^{\infty}({\bf R}^n,G)$
(where $G$ is an arbitary matrix Lie group) and prove that it is impossible to define 
a finite
translationally invariant Borel measure on it 
(provided that it is endowed with
a suitable topology).

The local gauge group $G_0=C^{\infty}({\bf R}^n,G)$ is the group of all smooth 
mappings from ${\bf R}^n$ to $G$. For a generic element $\omega(x^1,\cdots x^n)$ of
$G_0$, the quantity $\omega^{-1}(0)\partial_1\omega(0)$(here
by $\partial_1\omega(0)$ we certainly mean $\frac{\partial}{\partial
x^1}\omega(x)|_{x=0}$) belongs to the Lie algebra $g$ of $G$. For
$\xi=\sum_{i=1}^{m}u_i\xi_i \in g$(where $\{ \xi_1 \cdots \xi_m \}$  is a
basis of $g$) we define $f(\xi)=e^{-\sum_{i=1}^{m}u_{i}^2}$ and consider the
following bounded continuous function of $\omega$ (in some suitable topology
of the group $G_0$): \begin{equation}
F[\omega]=f(\omega^{-1}(0)\partial_1\omega(0)) \end{equation}

In the following we shall prove that it is impossible to define a {\it finite}
(left or right)
translationally invariant measure on $G_0$ by using the method of {\it reduction to
absurdity}. Suppose we have defined a measure $D\mu(\omega)$ which is right 
translationally invariant. Then we have the
following equality:
\begin{equation}
\int_{G_0}D\mu(\omega)F[\omega]= \int_{G_0}D\mu(\omega)F[\omega\omega_0],
\forall \omega_0 \in G_0
\end{equation}
From Eq(1) we have
\begin{eqnarray}
&\int_{G_0}&D\mu(\omega)f(\omega^{-1}(0)\partial_1\omega(0)) \nonumber \\
&=&\int_{G_0}f(\omega_0^{-1}(0)\omega^{-1}(0)\partial_1\omega(0)\omega_0(0)
+\omega_0^{-1}(0)\partial_1\omega_0(0))
, \forall \omega_0 \in G_0
\end{eqnarray}
In Eq(3) we set $\omega_0(x^1,\cdots x^n)=e^{\xi x^1}$, where $\xi$ is an arbitrary 
element of the Lie algebra $g$ and get
\begin{equation}
\int_{G_0}D\mu(\omega)f(\omega^{-1}(0)\partial_1\omega(0))=\int_{G_0}D\mu(\omega)f(
\omega^{-1}(0)\partial_1\omega(0) +\xi), \forall 
\xi \in g
\end{equation}
Let us consider the continuous function $f(\omega^{-1}(0)\partial_1\omega(0)+\xi)$ 
defined on $
G_0\times
g$ and observe that ($dm(\xi)=du^1\cdots du^m$
is the Lebesgue measure on $g$)
\begin{eqnarray}
\int_{G_0}D\mu(\omega)\int_g dm(\xi)|f(\omega^{-1}(0)\partial_1\omega(0)+\xi)|&=& 
\int_{G_0}D\mu(\omega)\int_g dm(\xi)f(\omega^{-1}(0)\partial_1\omega(0)+\xi) 
\nonumber \\
&=& \int_{G_0}D\mu(\omega)\int_g dm(\xi)f(\xi) \nonumber \\
&=& \pi^{n/2}\mu(G_0)
\end{eqnarray}
where we have used the translational invariance of the Lebesgue measure
$dm(\xi)$. According to Fubini
theorem another iterated integral also exists and is equal to the integral in Eq(5):
\begin{equation}
\int_g d m(\xi)\int_{G_0}
D\mu(\omega)f(\omega^{-1}(0)\partial_1\omega(0)+\xi)= \pi^{n/2}\mu(G_0)
\end{equation}
However, according to Eq(4), which expresses the right translational invariace
of $D\mu(\omega)$, we ought to have
\begin{equation}
\int_g dm(\xi)\int_{G_0} d\mu(\omega)f(\omega^{-1}(0)\partial_1\omega(0)+\xi)=+\infty
\end{equation}
(we have $\int_{G_0}D\mu(\omega)f(\omega^{-1}(0)\partial_1\omega(0)) >0$).
Thus we get a contradiction, which shows that a {\it finite} right translationally 
invariant
measure cannot be defined on $G_0$. By
considering the function $f(\omega(0)\partial_1\omega^{-1}(0))$ we can similarly 
show that a {\it finite} left translationally invariant measure cannot be defined
on $G_0$ either.

The result proved above does not put essential constraints to the Faddeev-Popov
formalism. It only requires that the volume of the local gauge group not
be finite, which is a commonly-taken understanding. It also does
not affect the construction of the gauge invariant ground state wave-functional
described in \cite{Kogan}. However the construction
of the projection operator for the Gauss law described in \cite{Huang} does have
difficulties. In this reference the invariant measure is formally taken to be
$D\mu(\omega)=\prod_x d\mu(\omega(x))$ where $d\mu(\omega(x))$ is the usual
invariant measure of group $G$ attached to the point $x$ normalized to be
$\int d\mu(\omega(x))=1$. However, the formal product of the measure
$d\mu(\omega(x))$ over the whole space ${\bf R}^n$ is not a measure in the
strict mathematical sense, and a finite invariant measure does not exist on
the local gauge group.


\begin{references}
\bibitem{Faddeev} L.Faddeev and V.N.Popov, Phys.Lett. B25(1967),29.
\bibitem{Huang} Kerson Huang, {\it Quarks, Leptons and Gauge Fields} 2nd ed.
World Scientific, 1992, sect.8.8.
\bibitem{Kogan} I.I.Kogan and A.Kovner, Phys.Rev.D51(1995),1948.
\bibitem{Halmos} P.R.Halmos, {\it Measure Theory}, Graduate Texts in Mathematics
18, Springer Verlag, 1974.
\bibitem{Yamasaki} Y.Yamasaki, {\it Measures on Infinite Dimensional Spaces},
World Scientific, 1985. 
\bibitem{SCW} W.M.Sun, Xiang-Song Chen and Fan Wang, hep-th/0105150 
\end{references}
\end{document}